\documentclass[prd,aps,showpacs,nofootinbib,preprint,eqsecnum]{revtex4}
\usepackage{graphicx}
\usepackage[english]{babel}
\usepackage{amsmath}
\usepackage{amssymb}
\usepackage{amsfonts}
\usepackage{latexsym}
\usepackage{color,amsxtra}
\usepackage{epsf}
\usepackage{enumerate}
\usepackage{hhline}
\usepackage{array}
\usepackage{tabularx}
\usepackage{subfigure}
\usepackage{fancyhdr}
\usepackage{mathrsfs}
\newcommand{\be}{\begin{equation}}
\newcommand{\ee}{\end{equation}}
\newcommand{\bea}{\begin{eqnarray}}
\newcommand{\eea}{\end{eqnarray}}
\newcommand{\beaa}{\begin{eqnarray*}}
\newcommand{\eeaa}{\end{eqnarray*}}

\newcommand{\Eqn}[1]{&\hspace{-0.2em}#1\hspace{-0.2em}&}

\def\be{\begin{equation}}
\def\ee{\end{equation}}
\def\bea{\begin{eqnarray}}
\def\eea{\end{eqnarray}}

\begin{document}

\title{Thermodynamic properties of modified gravity theories}

\author{Kazuharu Bamba}
\affiliation{
Division of Human Support System, Faculty of Symbiotic Systems Science, Fukushima University, Fukushima 960-1296, Japan}


\begin{abstract} 
We review thermodynamic properties of modified gravity theories 
such as $F(R)$ gravity and $f(T)$ gravity, where $R$ is the scalar curvature and $T$ is the torsion scalar in teleparallelism. 
In particular, we explore the equivalence between the equations of motion for modified gravity theories and the Clausius relation in thermodynamics. 
In addition, thermodynamics of the cosmological apparent horizon is 
investigated in $f(T)$ gravity. We show both equilibrium and non-equilibrium 
descriptions of thermodynamics. 
It is demonstrated that the second law of thermodynamics in the universe can be met when the temperature of the outside of the apparent horizon is 
equivalent to that of the inside of it.
\end{abstract}

\hspace{13.0cm} FU-PCG-10

\maketitle

\def\thesection{\Roman{section}}
\def\theequation{\Roman{section}.\arabic{equation}}

\section{Introduction}

According to the various cosmological observations including 
Type Ia Supernovae~\cite{Perlmutter:1998np, Riess:1998cb}, 
cosmic microwave background (CMB) radiation~\cite{Ade:2015xua, Ade:2015lrj, Ade:2014xna, Ade:2015tva, Array:2015xqh, Komatsu:2010fb, Hinshaw:2012aka}, 
large scale structure~\cite{Tegmark:2003ud, Seljak:2004xh}, 
baryon acoustic oscillations (BAO)~\cite{Eisenstein:2005su}, 
and weak lensing~\cite{Jain:2003tba}, 
the current expansion of the universe is accelerating. 
There exist two main procedures to account for the late-time cosmic 
acceleration. 
One is the introduction of the so-called ``dark energy'' 
in general relativity, and the other is the modification of 
gravitation (for reviews on dark energy problems and modified gravity theories, see, for instance,~\cite{Nojiri:2010wj, Nojiri:2006ri, Book-Capozziello-Faraoni, Sotiriou:2008rp, Capozziello:2011et, delaCruzDombriz:2012xy, Bamba:2012cp, Joyce:2014kja, Koyama:2015vza, Bamba:2013iga, Bamba:2014eea, Bamba:2015uma}). 

It has been suggested by black hole thermodynamics~\cite{Bardeen:1973gs, Bekenstein:1973ur, Hawking:1974sw, Gibbons:1977mu} 
with the black hole entropy~\cite{Bekenstein:1973ur} and a Hawking
temperature~\cite{Hawking:1974sw} 
that gravitation has a fundamental connection to thermodynamics 
(for reviews, see, e.g.,~\cite{Padmanabhan:2009ey, Padmanabhan:2009jb, Padmanabhan:2009vy, Padmanabhan:2015pza}). 
Indeed, in general relativity, 
the Einstein equation has been derived from 
the Clausius relation in thermodynamics by taking into account 
the fact that the entropy is proportional to the horizon 
area in Ref.~\cite{Jacobson:1995ab}. 
Furthermore, the investigations proposed in Ref.~\cite{Jacobson:1995ab} 
have been applied to extended theories of gravitation~\cite{Elizalde:2008pv, Wu:2009np, Yokokura:2011za, Brustein:2009hy}. 
It is known that when the gravitational field equation is derived with the procedure in Ref.~\cite{Jacobson:1995ab} in $F(R)$ gravity~\cite{Buchdahl:1983zz, Capozziello:2003tk, Carroll:2003wy, Nojiri:2003ft}, 
a treatment of a non-equilibrium thermodynamics would be necessary~\cite{Eling:2006aw}. It is important to note that the anti-evaporation of black holes 
have been considered in Refs.~\cite{Nojiri:2013su, Sebastiani:2013fsa, Nojiri:2014jqa, Katsuragawa:2014hda}. 

As a gravity theory alternative to general relativity, 
there is the so-called teleparallelism in which 
the Weitzenb\"{o}ck connection is used. 
In this theory, curvature does not exist, but 
torsion does. This is different from general relativity 
described by the Levi-Civita connection~\cite{Hehl:1976kj, Hayashi:1979qx, Flanagan:2007dc, Garecki:2010jj, Bamba:2015jqa}. 
The Lagrangian density for teleparallelism is written 
by the torsion scalar $T$. 
This has been extended to a function of $T$, namely, 
it is called $f(T)$ gravity, 
to explain not only inflation~\cite{Ferraro:2006jd, Ferraro:2008ey, Bamba:2014zra} but also the late-time accelerated expansion of the universe~\cite{Bengochea:2008gz, Linder:2010py, Bamba:2010wb, Bamba:2010iw, Geng:2011aj}. Such an approach is the same as $F(R)$ gravity. 
It is important to remark that there is no local Lorentz invariance in $f(T)$ gravity~\cite{Li:2010cg, Sotiriou:2010mv} and there are several discussions 
on the related points~\cite{Ferraro:2011us, Li:2011rn, I1, I2, I3, I4, Bahamonde:2015zma, Bamba:2013jqa, Bamba:2013ooa}. 
Moreover, finite-time future singularities in $f(T)$ gravity has been 
studied in Ref.~\cite{BMNO}. 
The first law of thermodynamics in $f(T)$ gravity has been examined 
in Ref.~\cite{Miao:2011ki}, in which the different procedure 
in Ref.~\cite{BG-2} has been used. 

In this paper, we review the main results in Refs.~\cite{BGNO, BG-2}. 
First, we investigate the equivalence of 
the gravitational field equation to the Clausius relation in thermodynamics 
in modified gravity theories. Particularly, we consider (a) $F(R)$ gravity, 
(b) the scalar-Gauss-Bonnet gravity inspired by (super)string theories 
(e.g., see~\cite{Nojiri:2005vv}), 
(c) $F(\mathcal{G})$ gravity~\cite{Nojiri:2005jg}, 
where $\mathcal{G}$ is the Gauss-Bonnet invariant, 
and
(d) the non-local gravity~\cite{Deser:2007jk, Nojiri:2007uq}. 
In addition, we discuss how to relate 
the expression of the entropy and the contribution of both matter and modification of gravity to the expression of the energy flux (heat). 
Next, for $f(T)$ gravity, we study thermodynamics of the apparent horizon. 
We examine both the non-equilibrium and equilibrium descriptions of thermodynamics. The dual equilibrium/non-equilibrium formulation in $f(T)$ gravity is found as in $F(R)$ gravity~\cite{BG-1}. 
We show that if the temperature of the outside of the apparent horizon is 
same as that of the inside of it, for the universe, the second law of thermodynamics can be satisfied. 
We use units of $k_\mathrm{B} = c = \hbar = 1$ and express the
gravitational constant $8 \pi G$ by
${\kappa}^2 \equiv 8\pi/{M_{\mathrm{Pl}}}^2$ 
with the Planck mass of $M_{\mathrm{Pl}} = G^{-1/2} = 1.2 \times 
10^{19}$\,\,GeV. 

The organization of the paper is the following. 
In Sec.~II, we explain the equivalence between the gravitational equations in modified gravity theories and the Clausius relation in thermodynamics. 
In Sec.~III, we consider thermodynamics of the apparent horizon in $f(T)$ gravity and present not only the non-equilibrium description but also the equilibrium one of thermodynamics. 
Conclusions are presented in Sec.~IV. 

\section{Equivalence between modified gravity equation and the Clausius relation} 

In this section, we review that in modified gravity theories, 
the gravitational field equation is equivalent to the Clausius relation in thermodynamics. 
We also consider the relation between the representation of the entropy and the contribution of not only matter but also modification of gravity 
to the expression of the energy flux (heat). 

\subsection{Formulations} 

The Clausius relation in thermodynamics 
is represented as~\cite{Jacobson:1995ab} $\delta S = \delta Q/T$ with 
$S$ the entropy, $Q$ the heat, and $T$ the temperature, 
where $\delta Q$ is considered to be the energy flux via the 
local Rindler horizon $\mathcal{H}$ at a free-falling local observer $p_0$. 
The expression of $\delta Q$ is 
$\delta Q = \int_{\mathcal{H}} T_{\mu\nu} \chi^{\mu} d \Sigma^{\nu}$. 
Here, $\chi^\mu$ is an approximate local boost Killing field 
future directed to the past of $p_0$, 
$T_{\mu\nu}$ is the energy-momentum tensor of all the matters, and 
the integration is over a pencil of generators of $\mathcal{H}$ at $p_0$. 
Furthermore, we have $d \Sigma^{\mu} = K^{\mu} d\lambda dA$. 
Here, $K^\mu$ is an approximate Killing field which generates boost at $p_0$ and vanishes at $p_0$. This is taken as the future pointing to the inside past of $p_0$. 
Moreover, $dA$ is the cross section area element of $\mathcal{H}$. 
On the other hand, $T$ is regarded as 
the Unruh temperature~\cite{Unruh:1976db}, given by
$T=k/\left(2\pi\right)$, where $k$ is the acceleration of the Killing orbit. 
On the Killing orbit, the norm of $\chi^{\mu}$ becomes unity 
when $K$ is a tangent vector to the generators of $\mathcal{H}$ with
an affine parameter $\lambda$. At $p_0$, we find $\lambda = 0$.

The formulation has been developed also in 
$F(R)$ gravity~\cite{Elizalde:2008pv, Eling:2006aw} 
and in extended gravity theories~\cite{Brustein:2009hy, Parikh:2009qs}. 
The first generalization of  to 
In the Lanczos-Lovelock gravity, 
the relation of the gravitational field equations to thermodynamics 
has been generalized in Refs.~\cite{Padmanabhan:2006ag, Paranjape:2006ca, Kothawala:2009kc}. 
In addition, for the Lanczos-Lovelock gravity, 
the entropy functional approach has been explored in Ref.~\cite{Padmanabhan:2007xy}. 
Except for a four-divergence, 
the procedure in Refs.~\cite{Brustein:2009hy, Wu:2008rp} 
is equivalent to this entropy functional approach. 
The connection of the entropy functional approach with diffeomorphism 
invariance has been indicated in Ref.~\cite{Padmanabhan:2009ry}. 
If a particular expression is introduced as entropy, 
in principle, all of the diffeomorphism invariant theories can obtain an entropic derivation. 

The entropy $S$ can be defined as~\cite{Wald:1993nt} 
\begin{equation}
S = - \frac{2}{T} \oint_{\partial\mathcal{H}}
S^{\mu\rho\nu\sigma} \hat\epsilon_{\mu\rho} \epsilon_{\nu\sigma}\,,
\label{eq:2.1}
\end{equation}
where 
\begin{equation}
S^{\mu\rho\nu\sigma} \equiv \frac{1}{\sqrt{-g}} \frac{\delta I}{\delta R_{\mu\rho\nu\sigma} }\,, 
\label{eq:2.2}
\end{equation}
with $I$ the action. 
This definition is used in Refs.~\cite{Elizalde:2008pv, Brustein:2009hy}. 
In Eq.~(\ref{eq:2.1}), the integration is over the surface which encloses the
volume $\mathcal{H}$, 
$\epsilon^{\mu\nu}$ is a 2-dimensional volume form, and 
$\hat\epsilon^{\mu\nu}$ is expressed as 
$\hat\epsilon^{\mu\nu} = \nabla^\mu \tilde\chi^\nu
= \epsilon^{\mu\nu}/ \bar\epsilon$, 
where $\tilde\chi^\nu = \chi^\nu/k$ and 
$\bar\epsilon$ is the area element of the cross section of the horizon. 
As a result, we acquire~\cite{Brustein:2009hy} 
\begin{equation}
T^{\sigma\nu} = 2 \left[ -2 \nabla_\mu \nabla_\rho S^{\mu\sigma\nu\rho} +
S^{\mu\rho\tau\sigma} R_{\mu\rho\tau}^{\ \ \ \nu} \right] + g^{\sigma\nu}
\Phi\,, 
\label{eq:2.3}
\end{equation}
where the conservation law or the Bianchi identity leads to $\Phi$. 
The scalar curvature, the Ricci tensor, and the Riemann tensor are defined as 
$R \equiv g^{\mu\nu}R_{\mu\nu}$, 
$R_{\mu\nu} \equiv R^\lambda_{\ \mu\lambda\nu}$, 
and 
$R^\lambda_{\ \mu\rho\nu} \equiv 
-\Gamma^\lambda_{\mu\rho,\nu}
+ \Gamma^\lambda_{\mu\nu,\rho}
- \Gamma^\eta_{\mu\rho}\Gamma^\lambda_{\nu\eta}
+ \Gamma^\eta_{\mu\nu}\Gamma^\lambda_{\rho\eta}$ 
with $\Gamma^\lambda_{\mu\rho,\nu}$ the connection, 
respectively. 
We note that the sign of the first term in the right-hand side (r.h.s.) 
of Eq.~(\ref{eq:2.3}) is different from that in Ref.~\cite{Brustein:2009hy} 
because the definition of the Riemann tensor is different.  
Thus, in Ref.~\cite{Brustein:2009hy}, it has been shown that 
the gravitational equations of motion are equivalent to the 
fundamental thermodynamic relation in generalized gravity theories. 

The energy flux $\delta Q$ is described with 
the energy-momentum tensor $T_{\mu\nu}$ of all the matters. 
The entropy $S$ is given in Eq.~(\ref{eq:2.1}) with Eq.~(\ref{eq:2.2}) 
in a gravity theory. 
By the Clausius relation $\delta S = \delta Q/T$, 
$\delta Q$ is related to $S$. 
We can find Eq.~(\ref{eq:2.3}) by combining 
the relation $\delta Q = \int_{\mathcal{H}} T_{\mu\nu} \chi^{\mu} d \Sigma^{\nu}$ and Eq.~(\ref{eq:2.1}) with the Clausius relation. 
Consequently, Eq.~(\ref{eq:2.3}) is 
the equation of motion in the gravity theory, 
through which the matter and gravity are related with each other.

\subsection{Modified gravity theories} 

In modified gravity theories, 
we investigate the representations of $\nabla_\mu \nabla_\rho S^{\mu\sigma\nu\rho}$, $S^{\mu\rho\tau\sigma} R_{\mu\rho\tau}^{\ \ \ \nu}$, and $\Phi$, which are components on the r.h.s. of Eq.~(\ref{eq:2.3}), 
to clearly illustrate the equations of motion from
the Clausius relation in thermodynamics. 
This procedure is an extension of the method in general relativity, 
proposed by Jacobson~\cite{Jacobson:1995ab} in order to explore 
the Einstein equation as a thermodynamic equation of state. 
Especially, we examine 
(a) $F(R)$ gravity, 
(b) the scalar-Gauss-Bonnet gravity, 
(c) $F(\mathcal{G})$ gravity~\cite{Nojiri:2005jg}, 
and
(d) the non-local gravity~\cite{Deser:2007jk, Nojiri:2007uq}. 
Here, the Gauss-Bonnet invariant is given by 
$\mathcal{G} \equiv R^2 -4R_{\mu\nu}R^{\mu\nu} +
R_{\mu\nu\rho\sigma}R^{\mu\nu\rho\sigma}$ 
and $F(\mathcal{G})$ is a function of $\mathcal{G}$. 
With Eq.~(\ref{eq:2.3}), 
it is demonstrated that in these gravity theories, 
the equations of motion are equivalent to the Clausius relation in thermodynamics. 
We mention that in Ref.~\cite{Eling:2006aw}, 
the equations of motion in $F(R)$ gravity have been studied 
in non-equilibrium thermodynamics, whereas 
in Ref.~\cite{Elizalde:2008pv}, they have been explored in 
equilibrium thermodynamics with the concept of 
``local-boost-invariance''~\cite{Iyer:1994ys}. 
We here generalize the Jacobson's approach considered in Ref.~\cite{Brustein:2009hy} to $F(R)$ gravity. 

We explore the following action for modified gravity theories 
\begin{equation}
I = \int d^4 x \sqrt{-g} \left[ \frac{\mathcal{F}(R,\phi,X,\mathcal{G})}
{2\kappa^2} +
{\mathcal{L}}_{\mathrm{matter}} \right]\,. 
\label{eq:2.4}
\end{equation}
Here, $g$ is the determinant of the metric $g_{\mu\nu}$, 
$\phi$ is a scalar field (for example, a dilaton for string theories), 
$X \equiv -\left(1/2\right) g^{\mu\nu} {\nabla}_{\mu}\phi {\nabla}_{\nu}\phi$
is a kinetic term of $\phi$, 
where ${\nabla}_{\mu}$ is the covariant derivative, 
$\mathcal{F}(R,\phi,X,\mathcal{G})$ is an arbitrary function in terms of
$R$, $\phi$, $X$ and $\mathcal{G}$, 
and 
${\mathcal{L}}_{\mathrm{matter}}$ is the Lagrangian of matter. 

It follows from the action in Eq.~(\ref{eq:2.4}) that the gravitational field equation is given by 
\begin{eqnarray}
&& 
{\mathcal{F}}_{,R} \left( R_{\mu\nu}-\frac{1}{2}R g_{\mu\nu}\right)
= \kappa^2 T^{(\mathrm{matter})}_{\mu \nu}
+\frac{1}{2}g_{\mu\nu} \left(\mathcal{F}-{\mathcal{F}}_{,R}R\right) 
+{\nabla}_{\mu}{\nabla}_{\nu}
{\mathcal{F}}_{,R} -g_{\mu\nu} \Box {\mathcal{F}}_{,R} 
\nonumber \\
&& 
{}+\frac{1}{2} {\mathcal{F}}_{,X} \partial_{\mu} \phi \partial_{\nu} \phi 
+\left(-2RR_{\mu\nu} +4R_{\mu\rho}R_{\nu}{}^{\rho} 
-2R_{\mu}{}^{\rho\sigma\tau}R_{\nu\rho\sigma\tau} 
+4g^{\alpha\rho}g^{\beta\sigma}R_{\mu\alpha\nu\beta}R_{\rho\sigma}
\right) {\mathcal{F}}_{,\mathcal{G}} 
\nonumber \\
&& 
{}+2\left({\nabla}_{\mu}{\nabla}_{\nu}{\mathcal{F}}_{,\mathcal{G}}\right)R 
-2g_{\mu \nu}\left(\Box {\mathcal{F}}_{,\mathcal{G}}\right)R
+4\left(\Box {\mathcal{F}}_{,\mathcal{G}}\right)R_{\mu \nu}
-4\left({\nabla}_{\rho}{\nabla}_{\mu}{\mathcal{F}}_{,\mathcal{G}}\right)
R_{\nu}{}^{\rho} 
-4\left({\nabla}_{\rho}{\nabla}_{\nu}{\mathcal{F}}_{,\mathcal{G}}\right)
R_{\mu}{}^{\rho}
\nonumber \\
&& 
{}+4g_{\mu \nu}\left({\nabla}_{\rho}{\nabla}_{\sigma}
{\mathcal{F}}_{,\mathcal{G}}\right)R^{\rho\sigma} 
-4\left({\nabla}_{\rho}{\nabla}_{\sigma}{\mathcal{F}}_{,\mathcal{G}}\right)
g^{\alpha\rho}g^{\beta\sigma}R_{\mu\alpha\nu\beta}\,. 
\label{eq:2.5} 
\end{eqnarray}
with
\begin{eqnarray}
{\mathcal{F}}_{,R} \equiv 
\frac{\partial \mathcal{F}(R,\phi,X,\mathcal{G})}{\partial R}\,, \quad
{\mathcal{F}}_{,X} \equiv 
\frac{\partial \mathcal{F}(R,\phi,X,\mathcal{G})}{\partial X}\,,
\nonumber \\ 
{\mathcal{F}}_{,\mathcal{G}} \equiv 
\frac{\partial \mathcal{F}(R,\phi,X,\mathcal{G})}
{\partial \mathcal{G}}\,, \quad
{\mathcal{F}}_{,\phi} \equiv 
\frac{\partial \mathcal{F}(R,\phi,X,\mathcal{G})}{\partial \phi}\,,
\label{eq:2.6}
\end{eqnarray}
where $\Box \equiv g^{\mu \nu} {\nabla}_{\mu} {\nabla}_{\nu}$
is the covariant d'Alembertian for a scalar field, and 
$T^{(\mathrm{matter})}_{\mu \nu}$ is the energy-momentum tensor of all the matters. 
Moreover, the equation of motion for $\phi$ is derived as 
\begin{equation}
{\mathcal{F}}_{,\phi} +\frac{1}{\sqrt{-g}}\partial_\mu
\left({\mathcal{F}}_{,X} \sqrt{-g} g^{\mu\nu}
\partial_\nu \phi\right) = 0\,,
\label{eq:2.7}
\end{equation}
%

\subsubsection{$F(R)$ gravity} 

In $F(R)$ gravity, the action is given by Eq.~(\ref{eq:2.4}) with 
\begin{equation}
\frac{\mathcal{F}(R,\phi,X,\mathcal{G})}{2\kappa^2} = F(R)\,.
\label{eq:2.8}
\end{equation}
In this case, we find 
\begin{eqnarray}
S^{\mu\nu\rho\sigma} \Eqn{=} \frac{F'(R)}{2}\left(g^{\mu\nu} g^{\rho\sigma}
 - g^{\mu\sigma} g^{\nu\rho} \right) \,,
\label{eq:2.9} \\
\nabla_\mu \nabla_\sigma S^{\mu\rho\nu\sigma}
\Eqn{=} \frac{1}{2} \left(\nabla^\nu \nabla^\rho - g^{\nu\rho} \Box \right)
F'(R)\,,
\label{eq:2.10} \\
S^{\mu\rho\tau\sigma} R_{\mu\rho\tau}^{\ \ \ \ \nu} \Eqn{=}
R^{\sigma\nu} F'(R)\,, 
\label{eq:2.11}
\end{eqnarray}
where $F'(R) \equiv dF(R)/dR$. 
Furthermore, the gravitational field equation reads 
\begin{equation}
0 = \frac{1}{2}g_{\mu\nu} F(R) - R_{\mu\nu}F'(R) + \nabla_\mu \nabla_\nu F'(R) 
- g_{\mu\nu}\Box F'(R) + \frac{1}{2} T^{(\mathrm{matter})}_{\mu\nu}\,. 
\label{eq:2.12}
\end{equation}
The comparison of Eq.~(\ref{eq:2.10}) with Eq.~(\ref{eq:2.3}) leads to 
\begin{eqnarray}
T_{\mu\nu} \Eqn{=} T^{(\mathrm{matter})}_{\mu \nu}\,,
\label{eq:2.13} \\
\Phi \Eqn{=} - F(R)\,, 
\label{eq:2.14}
\end{eqnarray}
where we have used using Eqs.~(\ref{eq:2.10}) and (\ref{eq:2.11}). 

\subsubsection{Scalar-Gauss-Bonnet gravity} 

The action for the scalar-Gauss-Bonnet gravity is expressed by 
Eq.~(\ref{eq:2.4}) with 
\begin{equation}
\frac{\mathcal{F}(R,\phi,X,\mathcal{G})}{2\kappa^2} = \frac{R}{2\kappa^2}
-\frac{\gamma}{2} g^{\mu\nu} {\partial}_{\mu}\phi {\partial}_{\nu}\phi
-V(\phi)+ f(\phi)\mathcal{G}\,, 
\label{eq:2.15}
\end{equation}
where $V(\phi)$ is the potential of $\phi$, 
$f(\phi)$ is a function of $\phi$, and 
$\gamma = \pm1$. 
When $\gamma =1$, $\phi$ is a canonical scalar field, 
while in the case that $\gamma = -1$ and 
there is no the Gauss-Bonnet invariant, 
$\phi$ is a phantom (non-canonical) scalar field.  
In this action, we obtain
\begin{eqnarray}
S^{\mu\rho\nu\sigma} \Eqn{=} \frac{1}{4\kappa^2}\left(g^{\mu\nu} g^{\rho\sigma} - g^{\mu\sigma} g^{\nu\rho} \right) 
+ f(\phi) \left\{ \left(g^{\mu\nu} g^{\rho\sigma} - g^{\mu\sigma} g^{\nu\rho}
\right) R
\right. \nonumber \\
&& \left. 
 - 2 \left(g^{\rho\sigma} R^{\mu\nu} - g^{\rho\nu} R^{\mu\sigma} - g^{\mu\sigma} R^{\rho\nu}
+ g^{\mu\nu} R^{\rho\sigma} \right) + 2 R ^{\mu\rho\nu\sigma} \right\}\,,
\label{eq:2.16} \\ 
\nabla_\mu \nabla_\sigma S^{\mu\rho\nu\sigma} 
\Eqn{=} \nabla_\mu \nabla_\sigma f(\phi) 
\left\{ \left(g^{\mu\nu} g^{\rho\sigma}
 - g^{\mu\sigma} g^{\nu\rho} \right) R
\right. \nonumber \\
&& \left. 
{}-2 \left(g^{\rho\sigma} R^{\mu\nu} - g^{\rho\nu} R^{\mu\sigma} - g^{\mu\sigma} R^{\rho\nu} \right)
+ 2 R ^{\mu\rho\nu\sigma} \right\} \,,
\label{eq:2.17} \\
S^{\mu\rho\nu\sigma}R_{\mu\rho\nu}^{\ \ \ \eta} 
\Eqn{=} \frac{1}{2\kappa^2} R^{\sigma\eta} + \frac{f(\phi) \mathcal{G}}{2} g^{\sigma\eta}\,. 
\label{eq:2.18}
\end{eqnarray}
In addition, the gravitational equation becomes 
\begin{eqnarray}
T^{(\mathrm{matter})}_{\mu \nu} \Eqn{=} 
\frac{1}{\kappa^2}\left(R_{\mu\nu} - \frac{1}{2}R g_{\mu\nu}\right) 
 - \gamma \left( \partial_{\mu} \phi \partial_{\nu} \phi
 - \frac{1}{2} g_{\mu\nu} \partial_{\rho} \phi \partial^{\rho} \phi \right)
+ g_{\mu\nu} V(\phi)
\nonumber \\
&& 
{}- 4 \left({\nabla}_{\mu}{\nabla}_{\nu}f(\phi)\right)R
 + 4 g_{\mu \nu}\left(\Box f(\phi)\right)R
 - 8 \left(\Box f(\phi)\right)R_{\mu \nu}
 + 8 \left({\nabla}_{\rho}{\nabla}_{\mu} f(\phi)\right)R_{\nu}{}^{\rho}
\nonumber \\
&&
{}+ 8 \left({\nabla}_{\rho}{\nabla}_{\nu} f(\phi)\right)R_{\mu}{}^{\rho} 
- 8 g_{\mu \nu}\left({\nabla}_{\rho}{\nabla}_{\sigma} f(\phi)\right) R^{\rho\sigma} 
\nonumber \\
&&
{}+ 8 \left({\nabla}_{\rho}{\nabla}_{\sigma} f(\phi)\right)
 g^{\alpha\rho}g^{\beta\sigma}R_{\mu\alpha\nu\beta}\,.
\label{eq:2.19}
\end{eqnarray}
We compare Eq.~(\ref{eq:2.19}) with Eq.~(\ref{eq:2.3}). 
Eventually, we acquire
\begin{eqnarray}
T_{\mu \nu} \Eqn{=}
T^{(\mathrm{matter})}_{\mu \nu} + \gamma \left( \partial_{\mu}
\phi \partial_{\nu} \phi
 - \frac{1}{2} g_{\mu\nu} \partial_{\rho} \phi \partial^{\rho} \phi \right) 
 - g_{\mu\nu} V(\phi)\,,
\label{eq:2.20} \\
\Phi \Eqn{=} \frac{R}{2\kappa^2} - f(\phi)\mathcal{G}\,.
\label{eq:2.21}
\end{eqnarray}
%

\subsubsection{$F(\mathcal{G})$ gravity} 

In $F(\mathcal{G})$ gravity, the action is represented by 
Eq.~(\ref{eq:2.1}) with~\cite{Nojiri:2005jg}
\begin{equation}
\frac{\mathcal{F}(R,\phi,X,\mathcal{G})}{2\kappa^2} = 
\frac{R}{2\kappa^2} + F(\mathcal{G}) \,. 
\label{eq:2.22}
\end{equation}
In this theory, we have 
\begin{eqnarray}
S^{\mu\rho\nu\sigma}
\Eqn{=}
\frac{1}{4\kappa^2}\left(g^{\mu\nu} g^{\rho\sigma} - g^{\mu\sigma} g^{\nu\rho}
\right) 
+ F'(\mathcal{G}) \left\{ \left(g^{\mu\nu} g^{\rho\sigma} - g^{\mu\sigma}
g^{\nu\rho} \right) R
\right. \nonumber \\
&& \left.
{}-2 \left(g^{\rho\sigma} R^{\mu\nu} - g^{\rho\nu} R^{\mu\sigma} - g^{\mu\sigma} R^{\rho\nu}
+ g^{\mu\nu} R^{\rho\sigma} \right)
+ 2 R ^{\mu\rho\nu\sigma} \right\}\,,
\label{eq:2.23} \\
\nabla_\mu \nabla_\sigma S^{\mu\rho\nu\sigma} \Eqn{=}
\nabla_\mu \nabla_\sigma F'(\mathcal{G}) \left\{ \left(g^{\mu\nu}
g^{\rho\sigma}
 - g^{\mu\sigma} g^{\nu\rho} \right) R
\right. \nonumber \\
&& \left.
{}- 2 \left(g^{\rho\sigma} R^{\mu\nu} - g^{\rho\nu} R^{\mu\sigma} - g^{\mu\sigma} R^{\rho\nu} \right)
+ 2 R ^{\mu\rho\nu\sigma} \right\} \,, 
\label{eq:2.24} \\
S^{\mu\rho\nu\sigma}R_{\mu\rho\nu}^{\ \ \ \eta}
\Eqn{=} \frac{1}{2\kappa^2} R^{\sigma\eta} +
\frac{F'(\mathcal{G}) \mathcal{G}}{2} g^{\sigma\eta}\,.
\label{eq:2.25} 
\end{eqnarray}
with $F'(\mathcal{G}) \equiv dF(\mathcal{G})/d\mathcal{G}$. 
Moreover, the gravitational field equation is given by 
\begin{eqnarray}
T^{(\mathrm{matter})}_{\mu\nu} \Eqn{=}
\frac{1}{\kappa^2}\left(R_{\mu\nu} - \frac{1}{2}R g_{\mu\nu}\right)
 - g_{\mu\nu}\left( F(\mathcal{G}) - \mathcal{G} F'(\mathcal{G}) \right) 
- 4 \left({\nabla}_{\mu}{\nabla}_{\nu}F'(\mathcal{G})\right)R
\nonumber \\
&& 
{}+ 4 g_{\mu \nu}\left(\Box F'(\mathcal{G}) \right)R 
 - 8 \left(\Box F'(\mathcal{G}) \right)R_{\mu \nu}
+ 8 \left({\nabla}_{\rho}{\nabla}_{\mu} F'(\mathcal{G})
\right)R_{\nu}{}^{\rho}
\nonumber \\
&& 
{}+ 8 \left({\nabla}_{\rho}{\nabla}_{\nu} F'(\mathcal{G}) \right)R_{\mu}{}^{\rho} - 8 g_{\mu \nu}\left({\nabla}_{\rho}{\nabla}_{\sigma} F'(\mathcal{G})
\right)R^{\rho\sigma} 
\nonumber \\
&& 
{}+ 8 \left({\nabla}_{\rho}{\nabla}_{\sigma} F'(\mathcal{G}) \right)
g^{\alpha\rho}g^{\beta\sigma}R_{\mu\alpha\nu\beta} \,.
\label{eq:2.26} 
\end{eqnarray}
By comparing Eq.~(\ref{eq:2.26}) with Eq.~(\ref{eq:2.3}) and 
using Eqs.~(\ref{eq:2.24}) and (\ref{eq:2.25}), 
we get 
\begin{eqnarray}
T_{\mu\nu} \Eqn{=} T^{(\mathrm{matter})}_{\mu \nu}\,,
\label{eq:2.27} \\
\Phi \Eqn{=} \frac{R}{2\kappa^2} - F(\mathcal{G})\,.
\label{eq:2.28} 
\end{eqnarray}
%

\subsubsection{Non-local gravity} 

In the non-local gravity, 
the action is described by Eq.~(\ref{eq:2.1}) with~\cite{Deser:2007jk, Nojiri:2007uq} 
\begin{equation}
\frac{\mathcal{F}(R,\phi,X,\mathcal{G})}{2\kappa^2} = 
\frac{1}{2\kappa^2}R\left(1 + \tilde{f}(\Box^{-1}R)\right) \,, 
\label{eq:2.29}
\end{equation}
where $\tilde{f}$ is a function of the argument. 
By introducing two scalar fields $\varphi$ and $\xi$, 
Eq.~(\ref{eq:2.29}) can be expressed as~\cite{Nojiri:2007uq}
\begin{equation}
\frac{\mathcal{F}(R,\phi,X,\mathcal{G})}{2\kappa^2} = 
\frac{1}{2\kappa^2}\left\{R\left(1 + \tilde{f}(\varphi)\right)
 - \partial_\mu \xi \partial^\mu \varphi - \xi R \right\} \,.
\label{eq:2.30}
\end{equation}
It follows from the expression of the action in Eq.~(\ref{eq:2.1}) with 
Eq.~(\ref{eq:2.30}) that 
\begin{eqnarray}
S^{\mu\rho\nu\sigma} 
\Eqn{=} \frac{1}{4\kappa^2}\left( 1 + \tilde{f}(\varphi) - \xi \right)
\left(g^{\mu\nu} g^{\rho\sigma} - g^{\mu\sigma} g^{\nu\rho} \right)\,,
\label{eq:2.31} \\
\nabla_\mu \nabla_\sigma S^{\mu\rho\nu\sigma}
\Eqn{=}
\frac{1}{4\kappa^2}\left(\nabla^\nu \nabla^\rho - g^{\nu\rho} \Box \right)
\left(\tilde{f}(\varphi) -\xi\right) \,, 
\label{eq:2.32} \\
S^{\mu\rho\nu\sigma}R_{\mu\rho\nu}^{\ \ \ \eta}
\Eqn{=} \frac{1}{2\kappa^2}\left( 1 + \tilde{f}(\varphi) - \xi\right)
R^{\sigma\eta}\,.
\label{eq:2.33} 
\end{eqnarray}
We also find the following gravitational equation
\begin{eqnarray}
T^{(\mathrm{matter})}_{\mu\nu} \Eqn{=} \frac{1}{\kappa^2} \left[
 - \frac{1}{2}g_{\mu\nu} \left\{ R\left(1 + \tilde{f}(\varphi) - \xi\right)
 - \partial_\rho \xi \partial^\rho \varphi \right\} 
+ R_{\mu\nu}\left(1 + \tilde{f}(\varphi) - \xi\right)
\right. \nonumber \\
&& \left.
{}- \frac{1}{2}\left(\partial_\mu \xi \partial_\nu \varphi
+ \partial_\mu \varphi \partial_\nu \xi \right) 
+ \left(g_{\mu\nu}\Box - \nabla_\mu \nabla_\nu\right)\left(
\tilde{f}(\varphi) - \xi\right) \right]\,.
\label{eq:2.34} 
\end{eqnarray}
In comparison of Eq.~(\ref{eq:2.34}) with Eq.~(\ref{eq:2.3}), 
with Eqs.~(\ref{eq:2.32}) and (\ref{eq:2.33}), we obtain
\begin{eqnarray}
T_{\mu\nu} \Eqn{=} T^{(\mathrm{matter})}_{\mu \nu}
+ \frac{1}{2\kappa^2}\bigl(\partial_\mu \xi \partial_\nu \varphi
+ \partial_\mu \varphi \partial_\nu \xi 
- g_{\mu\nu} \partial_\rho \xi 
\partial^\rho \varphi \bigr)\,,
\label{eq:2.35} \\
\Phi \Eqn{=} \frac{R}{2\kappa^2}\left( 1 + \tilde{f}(\varphi) -
\xi\right)\,.
\label{eq:2.36} 
\end{eqnarray}
Here, there is a possibility to select other separation 
into the parts of $T_{\mu\nu}$ and $\Phi$. 
Concretely, we can take 
\begin{eqnarray}
\hat{T}_{\mu\nu} \Eqn{=} T^{(\mathrm{matter})}_{\mu \nu}
+ \frac{1}{2\kappa^2}\left(\partial_\mu \xi \partial_\nu \varphi
+ \partial_\mu \varphi \partial_\nu \xi \right) \,, 
\label{eq:2.37} \\
\hat{\Phi} \Eqn{=} \frac{R}{2\kappa^2}\left( 1 + \tilde{f}(\varphi) -
\xi\right) - \partial_\rho \xi \partial^\rho \varphi \,, 
\label{eq:2.38} 
\end{eqnarray}
instead of combination of Eqs.~(\ref{eq:2.35}) and (\ref{eq:2.36}). 
Namely, we have included the last term in Eq.~(\ref{eq:2.35}) 
in the representation of $\hat{\Phi}$ in Eq.~(\ref{eq:2.38}). 

In general, $\Phi$ can be written as 
$\Phi = c_1 \left[ R/\left(2\kappa^2 \right) \right] + c_2 {\mathcal{L}}_{\mathrm{gravity}}$ 
with $c_1$ and $c_2$ constants, 
where $R/\left(2\kappa^2\right)$ describes general relativity and 
${\mathcal{L}}_{\mathrm{gravity}}$ is the Lagrangian of gravity. 
In $F(R)$ gravity, $(c_1, c_2) = (0, -1)$, and therefore 
${\mathcal{L}}_{\mathrm{gravity}} = F(R)$. 
In the scalar-Gauss-Bonnet gravity, $(c_1, c_2) = (2, -1)$, 
and hence 
${\mathcal{L}}_{\mathrm{gravity}} = R/\left(2\kappa^2\right) + 
f(\phi)\mathcal{G}$. 
In $F(\mathcal{G})$ gravity, $(c_1, c_2) = (2, -1)$, and thus 
${\mathcal{L}}_{\mathrm{gravity}} = 
R/\left(2\kappa^2\right) + F(\mathcal{G})$. 
In the non-local gravity, for Eq.~(\ref{eq:2.36}), 
$(c_1, c_2) = (0, 1)$, and accordingly 
${\mathcal{L}}_{\mathrm{gravity}} = 
\left[ R/\left(2\kappa^2\right) \right] 
\left(1 + \tilde{f}(\varphi) - \xi \right)$. 
When we take Eq.~(\ref{eq:2.38}), and not Eq.~(\ref{eq:2.36}), 
we have ${\mathcal{L}}_{\mathrm{gravity}} = 
\left[ R/\left(2\kappa^2\right) \right] 
\left(1 + \tilde{f}(\varphi) - \xi \right) - 
\partial_\rho \xi \partial^\rho \varphi$. 
This corresponds to Eq.~(\ref{eq:2.30}). 

As a result, we have confirmed that for modified gravity theories, 
the idea by Jacobson~\cite{Jacobson:1995ab} in general relativity 
that the Einstein equation can be considered to be a thermodynamic equation of state can be generalized. This point has been indicated in Ref.~\cite{Brustein:2009hy}, where gravity on a macroscopic scale is interpreted as 
a manifestation of thermodynamics in terms of the vacuum state of the quantum field theory.

\subsection{Representation of the entropy} 

In modified gravity theories, the gravitational field equation can 
be described as 
$T^{(\mathrm{matter})}_{\mu\nu} + T^{(\mathrm{MG})}_{\mu\nu}
= \left(1/\kappa^2 \right) G^{(\mathrm{GR})}_{\mu\nu}$ 
where $T^{(\mathrm{MG})}_{\mu\nu}$ is the energy-momentum tensor from 
the deviation of modified gravity theories from general relativity, 
and 
$G^{(\mathrm{GR})}_{\mu\nu} \equiv R_{\mu\nu} - \left(1/2\right)R g_{\mu\nu}$ 
is the Einstein tensor in general relativity. 
When the contribution from
$T^{(\mathrm{MG})}_{\mu\nu}$ is included in the representation of the energy 
flux (heat), 
the area law of the entropy is the same as the one in general relativity, 
and the contribution of the modification of gravity is involved 
in the entropy. 
Moreover, the gravitational field equation can be rewritten to  
$T^{(\mathrm{matter})}_{\mu\nu}
= \left(1/\kappa^2 \right) 
\left(G^{(\mathrm{GR})}_{\mu\nu} + G^{(\mathrm{MG})}_{\mu\nu} \right)$ 
with 
$G^{(\mathrm{MG})}_{\mu\nu} 
\equiv - \kappa^2 T^{(\mathrm{MG})}_{\mu\nu}$. 

On the other hand, 
the representation of the energy flux (heat) 
includes only the contribution of matter, 
the entropy $S$ would generally be described by a 
function of the area $A$ as $S=S(A)$, 
in which the parameters of modified gravity theories and/or curvatures, 
and so force, are included. 

In addition, there can be the mixture expression of 
the gravitational field equation as 
$T^{(\mathrm{matter})}_{\mu\nu} + \bar{T}^{(\mathrm{MG})}_{\mu\nu}
= \left(1/\kappa^2 \right) 
\left( G^{(\mathrm{GR})}_{\mu\nu} + 
\bar{G}^{(\mathrm{MG})}_{\mu\nu}\right)$, 
where 
$G^{(\mathrm{MG})}_{\mu\nu}
= - \kappa^2 T^{(\mathrm{MG})}_{\mu\nu} 
= \bar{G}^{(\mathrm{MG})}_{\mu\nu}
= - \kappa^2 \bar{T}^{(\mathrm{MG})}_{\mu\nu}$.
Hence, the contribution of both matter and 
a part of modification of gravity 
are involved in the entropy, 
and the representation of the entropy 
would be changed from that in general relativity. 
Therefore, it should be made clear 
whether the entropy contains the contribution 
of only matter or that of the modification of gravity 
in part as well as matter. 
Accordingly, we note that if any scalar field is included in the theory, 
the Wald's formula in Eq.~(\ref{eq:2.3}) cannot always be applicable. 

The point described above 
is related to the fact that in modified gravity theories, 
there can exist a description of equilibrium 
thermodynamics on the apparent horizon in the expanding universe 
because of a redefinition of an energy momentum tensor 
from the contribution of modification of gravity 
from which a local energy conservation is satisfied~\cite{Bamba:2009id, Bamba:2011jq}.

\section{Thermodynamics in $f(T)$ gravity} 

In this section, we review 
thermodynamics of the apparent horizon in $f(T)$ gravity. 
We consider the equilibrium description as well as the non-equilibrium description. We present the dual equilibrium/non-equilibrium formulation in $f(T)$ 
gravity, which is also obtained for $F(R)$ gravity in the Palatini formalism~\cite{BG-1}. 
It is shown that when the universe has the same temperature outside and inside 
the apparent horizon, the second law of thermodynamics can be met. 

The advantages of $f(T)$ gravity are summarized as follows. 
The cosmic acceleration, i.e., inflation and the late-time acceleration, 
can be realized in $f(T)$ gravity can realized. 
In addition, in $f(T)$ gravity, 
the gravitational field equation is second-order in derivatives, 
similar to that in general relativity, 
while in $F(R)$ gravity, we have 
the fourth-order gravitational field equation in derivatives. 
Accordingly, to investigate whether $f(T)$ gravity 
can be an alternative theory of gravitation to general relativity, 
it is significant to examine the first and second laws of thermodynamics 
in $f(T)$ gravity.

\subsection{Formulae in $f(T)$ gravity} 

Orthonormal tetrad components 
$e_A (x^{\mu})$ is used in the teleparallelism. 
Here, for the tangent space at each point $x^{\mu}$ of the manifold, 
an index $A$ runs over $0, 1, 2, 3$. 
Their relation to the metric $g^{\mu\nu}$ is given by 
$g_{\mu\nu}=\eta_{A B} e^A_\mu e^B_\nu$. 
Here, $\mu$ and $\nu$ are coordinate indices on the manifold 
and run over $0, 1, 2, 3$. 
Moreover, $e_A^\mu$ corresponds to the tangent vector of the manifold. 

The torsion tensor and the contorsion tensor 
are defined as 
$T^\rho_{\verb| |\mu\nu} \equiv e^\rho_A 
\left( \partial_\mu e^A_\nu - \partial_\nu e^A_\mu \right)$ 
and 
$K^{\mu\nu}_{\verb|  |\rho} \equiv 
-\left(1/2\right) 
\left(T^{\mu\nu}_{\verb|  |\rho} - T^{\nu \mu}_{\verb|  |\rho} - 
T_\rho^{\verb| |\mu\nu}\right)$, respectively. 
The Lagrangian density in the teleparallelism 
is given by the torsion scalar 
$T \equiv S_\rho^{\verb| |\mu\nu} T^\rho_{\verb| |\mu\nu}$, 
with
$S_\rho^{\verb| |\mu\nu} \equiv \left(1/2\right) 
\left(K^{\mu\nu}_{\verb|  |\rho}+\delta^\mu_\rho \ 
T^{\alpha \nu}_{\verb|  |\alpha}-\delta^\nu_\rho \ 
T^{\alpha \mu}_{\verb|  |\alpha}\right)$, 
although in general relativity, the Lagrangian density is written by 
the scalar curvature $R$. 

The action in $f(T)$ gravity is described as~\cite{Linder:2010py} 
\begin{equation}
I= \int d^4x |e| \left[ \frac{f(T)}{2{\kappa}^2} 
+{\mathcal{L}}_{\mathrm{M}} \right]\,. 
\label{eq:3.1}
\end{equation}
with $|e|= \det \left(e^A_\mu \right)=\sqrt{-g}$ and 
${\mathcal{L}}_{\mathrm{M}}$ the Lagrangian of matter. 
The variation of the action in Eq.~(\ref{eq:3.1}) with respect to 
the vierbein vector field $e_A^\mu$ leads to~\cite{Bengochea:2008gz} 
\begin{equation}
\frac{1}{e} \partial_\mu \left( eS_A^{\verb| |\mu\nu} \right) f^{\prime} 
-e_A^\lambda T^\rho_{\verb| |\mu \lambda} S_\rho^{\verb| |\nu\mu} 
f^{\prime} +S_A^{\verb| |\mu\nu} \partial_\mu \left(T\right) f^{\prime\prime} 
+\frac{1}{4} e_A^\nu f = \frac{{\kappa}^2}{2} e_A^\rho 
{T^{(\mathrm{M})}}_\rho^{\verb| |\nu}\,. 
\label{eq:3.2}
\end{equation}
Here, ${T^{(\mathrm{M})}}_\rho^{\verb| |\nu}$ 
is the energy-momentum tensor of all perfect fluids of matter, 
namely, radiation and non-relativistic matter, and 
the prime denotes the derivative with respect to $T$. 

We suppose the four-dimensional flat 
Friedmann-Lema\^{i}tre-Robertson-Walker (FLRW) 
space-time, in which the metric is expressed as 
$d s^2 = h_{\alpha \beta} d x^{\alpha} d x^{\beta}
+\tilde{r}^2 d \Omega^2$. 
Here, $\tilde{r}=a(t)r$, $x^0=t$ and $x^1=r$ with the two-dimensional 
metric $h_{\alpha \beta}={\rm diag}(1, -a^2(t))$, 
where $a(t)$ is the scale factor and 
$d \Omega^2$ is the metric of two-dimensional sphere which has unit radius. 
For this space-time, we have
$g_{\mu \nu}= \mathrm{diag} (1, -a^2, -a^2, -a^2)$. 
With the tetrad components $e^A_\mu = (1,a,a,a)$, we find 
the relation $T=-6H^2$ between $T$ and the Hubble parameter 
$H=\dot{a}/a$, where the dot means the time derivative.  

In the flat FLRW space-time, from Eq.~(\ref{eq:3.2}), 
the gravitational field equations read~\cite{Bengochea:2008gz, Linder:2010py} 
\begin{eqnarray}
H^2 \Eqn{=} \frac{1}{6 F} 
\left( {\kappa}^2 \rho_{\mathrm{M}} -\frac{f}{2} \right)\,, 
\label{eq:3.3} \\ 
\dot{H} 
\Eqn{=} 
-\frac{1}{ 4TF^{\prime} + 2F } 
\left( {\kappa}^2 P_{\mathrm{M}} -TF +\frac{f}{2} \right)\,. 
\label{eq:3.4}
\end{eqnarray}
Here, $F \equiv df/dT$, $F^{\prime} = dF/dT$, and 
$\rho_{\mathrm{M}}$ and $P_{\mathrm{M}}$ are 
the energy density and pressure of all perfect fluids of matter, 
respectively. 
The continuity equation for the perfect fluid is satisfied 
as $\dot{\rho}_{\mathrm{M}}+3H\left( \rho_{\mathrm{M}} + P_{\mathrm{M}} \right)
=0$.

\subsection{Description of non-equilibrium thermodynamics for $f(T)$ gravity} 

\subsubsection{First law of thermodynamics in non-equilibrium description}

We see that Eqs.~(\ref{eq:3.3}) and (\ref{eq:3.4}) can be 
represented as 
\begin{eqnarray}  
H^2 
\Eqn{=}
\frac{\kappa^2}{3F} \left( \hat{\rho}_{\mathrm{DE}}+
\rho_{\mathrm{M}} \right)\,, 
\label{eq:3.5} \\
\dot{H}
\Eqn{=}
-\frac{\kappa^2}{2F} \left( \hat{\rho}_{\mathrm{DE}}+\hat{P}_{\mathrm{DE}}
+\rho_{\mathrm{M}}+P_{\mathrm{M}} \right)\,,
\label{eq:3.6} \\
\hat{\rho}_{\mathrm{DE}} 
\Eqn{\equiv} 
\frac{1}{2\kappa^2}\left( FT - f \right)\,, 
\label{eq:3.7} \\
\hat{P}_{\mathrm{DE}} 
\Eqn{\equiv} 
\frac{1}{2\kappa^2} 
\left[ -\left( FT - f \right) + 4H\dot{F}
\right]\,. 
\label{eq:3.8}
\end{eqnarray} 
Here, $\hat{\rho}_{\mathrm{DE}}$ and $\hat{P}_{\mathrm{DE}}$ are 
the energy density and pressure of dark components, 
respectively, and they meet 
$\dot{\hat{\rho}}_{\mathrm{DE}}+3H \left( 
\hat{\rho}_{\mathrm{DE}}+\hat{P}_{\mathrm{DE}} \right)
=-\left[T/\left(2\kappa^2\right)\right] \dot{F}$, 
where the hat shows quantities in the non-equilibrium 
description of thermodynamics. 
It is found that the standard continuity equation 
cannot be satisfied because of $\dot{F}\neq 0$. 

The dynamical apparent horizon is determined by the relation
$h^{\alpha \beta} \partial_{\alpha} \tilde{r} \partial_{\beta} \tilde{r}=0$. 
It is suggested that 
according to the observational data of type Ia Supernovae, 
the generalized second law of thermodynamics can be satisfied 
for the apparent horizon, and not for the event horizon~\cite{Zhou:2007pz,Wang:2005pk}. 
In the flat FLRW universe, the radius $\tilde{r}_A$ of 
the apparent horizon is represented as 
$\tilde{r}_A= 1/H$. 
It follows from the time derivative of this relation that 
$-d\tilde{r}_A/\tilde{r}_A^3 = \dot{H}H dt$. 
By substituting Eq.~(\ref{eq:3.6}) into this relation, we acquire 
\begin{equation}
\frac{F}{4\pi G} d\tilde{r}_A=\tilde{r}_A^3 H
\left( \hat{\rho}_{\mathrm{t}}+\hat{P}_{\mathrm{t}}\right) dt\,, 
\label{eq:3.9}
\end{equation}
with 
$\hat{\rho}_{\mathrm{t}} \equiv \hat{\rho}_{\mathrm{DE}}+
\rho_{\mathrm{M}}$ and $\hat{P}_{\mathrm{t}} \equiv \hat{P}_{\mathrm{DE}}+
P_{\mathrm{M}}$ the total energy density and pressure of the universe, respectively. 

The Bekenstein-Hawking horizon entropy is expressed as 
$S=A/\left(4G\right)$ in general relativity. 
Here, $A=4\pi \tilde{r}_A^2$ is the area of the 
apparent horizon~\cite{Bardeen:1973gs, Bekenstein:1973ur, Hawking:1974sw, Gibbons:1977mu}. 
On the other hand, 
for modified gravity theories such as $f(R)$ gravity, 
the Wald entropy $\hat{S}$~\cite{Wald:1993nt, Iyer:1994ys}, 
which is a horizon entropy $\hat{S}$ associated with a Noether 
charge, is described by 
$\hat{S}=A/\left(4G_{\rm eff}\right)$. 
Here, $G_{\mathrm{eff}}=G/f'$ 
with $f' = d f(R)/dR$ 
is the effective gravitational coupling 
in $f(R)$ gravity~\cite{Brustein:2007jj}. 
The form of the Wald entropy in $f(R)$ gravity 
in the metric formalism~\cite{Wald:1993nt, Iyer:1994ys, Jacobson:1993vj, Cognola:2005de} is equivalent to that in the 
Palatini formalism~\cite{Vollick:2007fh}. 

{}From the investigations of the matter density perturbations, 
the effective gravitational coupling in $f(T)$ gravity 
becomes $G_{\mathrm{eff}}=G/F$~\cite{Zheng:2010am}, 
which is similar to that for $f(R)$ gravity. 
Moreover, 
with the method of the Wald's Noether charge~\cite{Wald:1993nt, Iyer:1994ys} 
and the related consequences in Refs.~\cite{Jacobson:1995ab, Brustein:2009hy, Eling:2006aw, Gu:2010wv, Miao:2011er}, 
it has been demonstrated in Ref.~\cite{Miao:2011ki} 
that if $F^{\prime}=dF(T)/dT = d^2 f(T)/d T^2$ is small, 
the black hole entropy is approximately equal to $FA/\left(4G\right)$ 
for $f(T)$ gravity. 
Thus, for $f(T)$ gravity, the Wald entropy can be taken as 
\begin{equation}
\hat{S}=\frac{FA}{4G}\,. 
\label{eq:3.10}
\end{equation}
In $f(T)$ gravity, there is no local Lorentz 
invariant~\cite{Li:2010cg, Sotiriou:2010mv}. 
This implies that there exist new degrees of freedom. 
At the background level, however, there is no new degrees of 
freedom, whereas at the level of the linear perturbation, 
only the constraint equations are met by the new vector
degree of freedom~\cite{Li:2010cg, Sotiriou:2010mv}. 
In the early universe, the new degrees of freedom seems not to contribute to physical observables directly~\cite{Wu:2011kh}. 
Therefore, it is considered that the new degrees of freedom will not influence on the entropy. 
With Eqs.~(\ref{eq:3.9}) and (\ref{eq:3.10}), we get 
\begin{equation}
\frac{1}{2\pi \tilde{r}_A} d\hat{S}=4\pi \tilde{r}_A^3 H
\left( \hat{\rho}_{\mathrm{t}}+\hat{P}_{\mathrm{t}} \right) dt +
\frac{\tilde{r}_A}{2G} dF\,. 
\label{eq:3.11}
\end{equation}

We have the Hawking temperature 
$T_{\mathrm{H}} = \left|\kappa_{\mathrm{sg}}\right|/\left(2\pi \right)$, 
which corresponds to the associated temperature of the apparent horizon, 
where $\kappa_{\mathrm{sg}} = \left[1/\left(2\sqrt{-h}\right)\right] \partial_\alpha \left( \sqrt{-h}h^{\alpha\beta} \partial_\beta \tilde{r} \right)$ 
with $h$ the determinant of the metric $h_{\alpha\beta}$ 
is the surface gravity~\cite{Cai:2005ra}, 
and it is written as
\begin{equation}
\kappa_{\mathrm{sg}} 
= -\frac{1}{\tilde{r}_A}
\left( 1-\frac{\dot{\tilde{r}}_A}{2H\tilde{r}_A} \right)
=-\frac{\tilde{r}_A}{2} \left( 2H^2+\dot{H} \right) 
= -\frac{2\pi G}{3F} \tilde{r}_A 
\left( \hat{\rho}_{\mathrm{t}}-3\hat{P}_{\mathrm{t}} \right)\,. 
\label{eq:3.12}
\end{equation} 
It is seen from Eq.~(\ref{eq:3.11}) that 
when the total equation of state (EoS) 
$w_{\mathrm{t}} \equiv \hat{P}_{\mathrm{t}}/\hat{\rho}_{\mathrm{t}}$ 
satisfies the condition $w_{\mathrm{t}} \le 1/3$, 
we have $\kappa_{\mathrm{sg}} \le 0$. Eventually, we obtain 
\begin{equation}
T_{\mathrm{H}}=\frac{1}{2\pi \tilde{r}_A}
\left( 1-\frac{\dot{\tilde{r}}_A}{2H\tilde{r}_A} \right)\,. 
\label{eq:3.13}
\end{equation}
The multiplication of the term $1-\dot{\tilde{r}}_A/(2H\tilde{r}_A)$ for 
Eq.~(\ref{eq:3.10}) leads to
\begin{equation}
T_{\mathrm{H}} d\hat{S} = 4\pi \tilde{r}_A^3 H \left(\hat{\rho}_{\mathrm{t}}+
\hat{P}_{\mathrm{t}} \right) dt 
-2\pi  \tilde{r}_A^2 \left(\hat{\rho}_{\mathrm{t}}+\hat{P}_{\mathrm{t}} 
\right) d \tilde{r}_A
+\frac{T_{\mathrm{H}}}{G}\pi \tilde{r}_A^2 dF\,. 
\label{eq:3.14}
\end{equation}

The Misner-Sharp energy~\cite{Misner:1964je, Bak:1999hd}
is defined by $E \equiv \tilde{r}_A/\left(2G\right)$ 
for general relativity. 
This can be extended to $\hat{E}=\tilde{r}_AF/\left(2G\right)$ 
because $G_{\mathrm{eff}}=G/F$~\cite{Zheng:2010am} 
for $f(T)$ gravity, 
similar to that for $f(R)$ gravity~\cite{Gong:2007md, Wu:2007se, Wu:2008ir} 
(see Refs.~\cite{Sakai:2001gh, Cai:2009qf} as related studies). 
By using this relation and $\tilde{r}_A= 1/H$, we find 
$\hat{E}=V \left[ 3F H^2/\left(8\pi G\right) \right] 
= V\hat{\rho}_{\mathrm{t}}$ 
with $V=4\pi \tilde{r}_A^3/3$ the volume inside 
the apparent horizon. Here, 
the last equality suggests that $\hat{E}$ is equivalent to the total intrinsic energy. 
It is seen that since $\hat{E} >0$, we find $F>0$. 
Hence, in $f(T)$ gravity, 
the effective gravitational coupling $G_\mathrm{eff} = G/F$ is 
positive, similarly to that in $f(R)$ gravity~\cite{Sotiriou:2008rp}. 
This condition is necessary for the graviton not to be a ghost from the quantum theoretical point of view~\cite{Starobinsky:2007hu}. 

{}From the continuity equations, we obtain 
\begin{equation}
d\hat{E} = -4\pi \tilde{r}_A^3 H \left(\hat{\rho}_{\mathrm{t}}
+\hat{P}_{\mathrm{t}} \right) dt 
+4\pi \tilde{r}_A^2  \hat{\rho}_{\mathrm{t}} 
d\tilde{r}_A+\frac{\tilde{r}_A }{2G} dF\,.
\label{eq:3.15}
\end{equation}
By combining Eqs.~(\ref{eq:3.14}) and (\ref{eq:3.15}), we have 
\begin{equation}
T_{\mathrm{H}} d\hat{S} = d\hat{E}+2\pi \tilde{r}_A^2 
\left(\hat{\rho}_{\mathrm{d}}+\rho_{\mathrm{f}}-\hat{P}_{\mathrm{d}}
-P_{\mathrm{f}}\right) d\tilde{r}_A 
+\frac{\tilde{r}_A}{2G} \left( 1+2\pi \tilde{r}_A T_{\mathrm{H}} \right) dF\,. 
\label{eq:3.16}
\end{equation}
We introduce 
the work density~\cite{Hayward:1997jp, Hayward:1998ee, Cai:2006rs} 
$\hat{W} \equiv -\left(1/2\right) \left( T^{(\mathrm{M})\alpha\beta}
h_{\alpha\beta} + \hat{T}^{(\mathrm{DE})\alpha\beta} h_{\alpha\beta} 
\right) = \left(1/2\right) \left(\hat{\rho}_{\mathrm{t}} 
-\hat{P}_{\mathrm{t}} \right)$, 
where $\hat{T}^{(\mathrm{DE})\alpha\beta}$ is 
the energy-momentum tensor of dark components. 
With the work density $\hat{W}$, we rewrite 
Eq.~(\ref{eq:3.16}) as 
\begin{equation}
T_{\mathrm{H}} d\hat{S}=-d\hat{E}+\hat{W} dV
+\frac{\tilde{r}_A}{2G} 
\left( 1+2\pi \tilde{r}_A T_{\mathrm{H}} \right) dF\,.
\label{eq:3.17}
\end{equation}
This relation is represented as 
$T_{\mathrm{H}} d\hat{S}+T_{\mathrm{H}} d_{i}\hat{S}=-d\hat{E}+\hat{W} dV$ 
with 
\begin{eqnarray}
d_{i}\hat{S} \Eqn{=} -\frac{1}{T_{\mathrm{H}}} \frac{\tilde{r}_A}{2G}
\left( 1+2\pi \tilde{r}_A T_{\mathrm{H}} \right) dF
=-\left( \frac{\hat{E}}{T_{\mathrm{H}}}+\hat{S} \right) 
\frac{dF}{F} 
\label{eq:3.18} \\ 
\Eqn{=} \frac{6\pi}{G} \frac{8HT+\dot{T}}
{T\left(4HT+\dot{T}\right)} dF\,. 
\label{eq:3.19}
\end{eqnarray}
Here, the term $d_{i}\hat{S}$ can be regarded as 
an entropy production term 
in the description of non-equilibrium thermodynamics. 
For $f(T)$ gravity, except for the case that $f(T)=T$, 
in which $F=1$ and $d_{i}\hat{S}=0$, 
$d_{i}\hat{S}$ in Eq.~(\ref{eq:3.18}) does not 
vanish. 
Thus, the first-law of equilibrium thermodynamics is satisfied.

\subsubsection{Second law of thermodynamics in non-equilibrium description}

For ordinary fluid dynamics in cosmology, 
the entropy is simply the fluid-entropy current, 
and therefore it is not related to the horizon entropy. 
In the flat FLRW background, 
the Bekenstein-Hawking horizon entropy of the apparent horizon 
is described by 
$S=A/\left(4G\right) = \pi/\left( G H^2 \right) \propto H^{-2}$. 
Here, we have used $A=4\pi \tilde{r}_A^2$ and $\tilde{r}_A= 1/H$ 
to derive the first equality. 
In modified gravity theories such as $F(R)$ gravity and $f(T)$ gravity, 
the phantom phase in which $\dot{H} > 0$ can exist. 
For this phase, 
$\dot{S} = -2\left[ \pi/\left( G H^3 \right) \right] \dot{H} < 0$. 
Hence, on the horizon entropy, 
the second law of thermodynamics cannot be met. 
As a consequence, 
modified gravity theories in which there exists the phantom phase 
cannot be an alternative theory of gravity to general relativity. 
Indeed, however,  
when we examine the entropy of the total energy of the 
horizon, namely, both the horizon entropy and 
the entropy of ordinary perfect fluids of matter, 
the total amount of the entropy always becomes large in time, 
and therefore the second law of thermodynamics can be satisfied. 
This has been demonstrated for $F(R)$ gravity~\cite{BG-1}. 
We explore this point for $f(T)$ gravity. 

The Gibbs equation in terms of all the matters and energy fluids 
is represented as 
\begin{equation}
T_{\mathrm{H}} d\hat{S}_{\mathrm{t}} = d\left( \hat{\rho}_{\mathrm{t}} V \right) +\hat{P}_{\mathrm{t}} dV 
= V d\hat{\rho}_{\mathrm{t}} + \left( \hat{\rho}_{\mathrm{t}} + \hat{P}_{\mathrm{t}} \right) dV\,, 
\label{eq:3.20}
\end{equation}
where $T_{\mathrm{H}}$ is the temperature of total energy inside the horizon and $\hat{S}_{\mathrm{t}}$ is the entropy of it. 
Here, we have supposed that the temperature of the outside 
of the apparent horizon is equal to that of the inside of it. 
The following relation is required 
in order for the second law of thermodynamics to be obeyed~\cite{Wu:2008ir}
\begin{equation}
\Xi \equiv 
\frac{d\hat{S}}{dt} + \frac{d\left( d_i \hat{S} \right)}{dt} 
+ \frac{d\hat{S}_{\mathrm{t}}}{dt} \geq 0\,. 
\label{eq:3.21}
\end{equation}
{}From the relation $T_{\mathrm{H}} d\hat{S}+T_{\mathrm{H}} d_{i}\hat{S}=-d\hat{E}+\hat{W} dV$ and Eqs.~(\ref{eq:3.5}) and (\ref{eq:3.20}), we get 
$\Xi = -\left[3/\left(4G\right)\right] \left[\left(\dot{T}^2 
F\right)/T^3\right]$. 
Here, $-T^3 = 216 H^6 >0$, and therefore the condition of $\Xi\geq 0$ 
is rewritten to 
$J \equiv \dot{T}^2 F = 144H^2 \dot{H}^2 F \geq 0$. 
Since $F>0$ owing to $\hat{E} >0$, 
this relation is always satisfied. 
As a result, in $f(T)$ gravity, 
the second law of thermodynamics can be met. 
The condition of $J \geq 0$ is independent of the sign of $\dot{H}$. 
This consequence is compatible with a phantom model with thermodynamics~\cite{Nojiri:2005sr}. Entropy in phantom models have been studied in Refs.~\cite{Brevik:2004sd, Nojiri:2004pf, Bilic:2008zk, Bamba:2009ay}. 

Here, the temperature of the apparent horizon, namely, 
the Hawking temperature $T_{\mathrm{H}} = |\kappa_{\mathrm{sg}}|/\left(2\pi\right)$ has been used as the physical temperature. 
It is seen from Eq.~(\ref{eq:3.12}) that 
this temperature depends on 
the energy-momentum tensor of the dark components 
coming from $f(T)$ gravity. 
The temperature of matters in the universe is $2.73$K of the CMB 
photons. In our investigations, only the case that 
the cosmic temperature inside the apparent horizon is the same as 
that of the horizon. 
Namely, the temperature of the apparent horizon is considered to be 
equivalent to the temperature of matters such as that of the CMB photons.

\subsection{Description of equilibrium thermodynamics for $f(T)$ gravity} 

In the non-equilibrium description, 
the non-vanishing entropy production term $d_i \hat{S}$ exists, 
and accordingly the continuity equation 
in terms of $\hat{\rho}_{\mathrm{DE}}$ in Eq.~(\ref{eq:3.7}) 
and $\hat{P}_{\mathrm{DE}}$ in Eq.~(\ref{eq:3.8}) 
cannot hold. 
In this section, we show that the entropy production term does not 
appear thanks to the redefinition of the energy density and pressure of dark components in order to satisfy the continuity equation. 
This is referred to as the equilibrium description of thermodynamics 
in $f(T)$ gravity. Such a procedure has been proposed in Refs.~\cite{Bamba:2009id, Bamba:2011jq}.

\subsubsection{First law of thermodynamics in equilibrium description}

We compare the gravitational field equations (\ref{eq:3.3}) and 
(\ref{eq:3.4}) with those in general relativity, 
given by 
\begin{eqnarray}
H^2 \Eqn{=} \frac{{\kappa}^2}{3} \left(\rho_{\mathrm{M}}+\rho_{\mathrm{DE}} 
\right)\,, 
\label{eq:3.22} \\ 
\dot{H}
\Eqn{=} -\frac{{\kappa}^2}{2} \left(\rho_{\mathrm{M}} + P_{\mathrm{M}} + 
\rho_{\mathrm{DE}} + P_{\mathrm{DE}} \right)\,, 
\label{eq:3.23} \\ 
\rho_{\mathrm{DE}} 
\Eqn{=} 
\frac{1}{2{\kappa}^2} 
\left( -T -f +2TF \right)\,, 
\label{eq:3.24} \\ 
P_{\mathrm{DE}} 
\Eqn{=} 
-\frac{1}{2{\kappa}^2} 
\left[ 
4\left(1 -F -2TF^{\prime} \right) \dot{H} 
+\left( -T -f +2TF \right)
\right]\,, 
\label{eq:3.25} 
\end{eqnarray}
where $\rho_{\mathrm{DE}}$ and $P_{\mathrm{DE}}$ are 
the energy density and pressure of dark components, 
respectively. 
The continuity equation $\dot{\rho}_{\mathrm{DE}}+3H \left( 
\rho_{\mathrm{DE}} + P_{\mathrm{DE}} \right) = 0$ is satisfied.

For the gravitational field equations (\ref{eq:3.22}) and (\ref{eq:3.23}), 
Eq.~(\ref{eq:3.9}) is described as 
\begin{equation} 
\frac{1}{4\pi G} d\tilde{r}_A=\tilde{r}_A^3 H
\left( \rho_{\mathrm{t}}+P_{\mathrm{t}} \right) dt\,,  
\label{eq:3.26}
\end{equation} 
where 
$\rho_{\mathrm{t}} \equiv \rho_{\mathrm{DE}}+
\rho_{\mathrm{M}}$ and $P_{\mathrm{t}} \equiv P_{\mathrm{DE}}+P_{\mathrm{M}}$. 
We introduce the horizon entropy $S=A/(4G)$. 
With Eq.~(\ref{eq:3.26}), we obtain 
\begin{equation} 
\frac{1}{2\pi \tilde{r}_A} dS=4\pi \tilde{r}_A^3 H
\left( \rho_{\mathrm{t}}+P_{\mathrm{t}} \right) dt \,. 
\label{eq:3.27}
\end{equation} 
By using the horizon temperature in Eq.~(\ref{eq:3.13}) and 
Eq.~(\ref{eq:3.27}), we acquire
\begin{equation} 
T_{\mathrm{H}} dS = 4\pi \tilde{r}_A^3 H \left(\rho_{\mathrm{t}}+P_{\mathrm{t}} \right) dt 
-2\pi  \tilde{r}_A^2 \left(\rho_{\mathrm{t}}+P_{\mathrm{t}} \right) 
d\tilde{r}_A\,. 
\label{eq:3.28}
\end{equation} 

The Misner-Sharp energy is defined as 
$E \equiv \tilde{r}_A/\left(2G\right) = V\rho_{\mathrm{t}}$. 
Therefore, we have
\begin{equation} 
dE=-4\pi \tilde{r}_A^3 H \left(\rho_{\mathrm{t}}+P_{\mathrm{t}} \right) dt 
+4\pi \tilde{r}_A^2 \rho_{\mathrm{t}} d\tilde{r}_A\,.  
\label{eq:3.29}
\end{equation} 
Here, the term proportional to $dF$ does not exist 
on the r.h.s. owing to the continuity equation. 
By combining Eqs.~(\ref{eq:3.28}) and (\ref{eq:3.29}), we obtain 
the equation describing the first law of 
equilibrium thermodynamics is found as 
$T_{\mathrm{H}} dS=-dE+W dV$, where 
the work density $W$ is represented as 
$W = \left(1/2\right) \left( \rho_{\mathrm{t}}-P_{\mathrm{t}} \right)$. 
As a consequence, through the redefinition of $\rho_{\mathrm{DE}}$ and $P_{\mathrm{DE}}$ in order for the continuity equation to hold, 
a description of equilibrium thermodynamics can be realized. 

Furthermore, from Eqs.~(\ref{eq:3.22}), (\ref{eq:3.23}), and (\ref{eq:3.27}), 
we get 
\begin{equation}
\dot{S} = 
8\pi^2 H \tilde{r}_A^4 \left(\rho_{\mathrm{t}}+P_{\mathrm{t}}\right) 
= \frac{6\pi}{G} \frac{\dot{T}}{T^2}\,. 
\label{eq:3.30}
\end{equation}
Hence, for the expanding universe, 
if the null energy condition 
$\rho_{\mathrm{t}}+P_{\mathrm{t}} \ge 0$ leading to $\dot{H} \leq 0$ 
is satisfied, the horizon entropy becomes large 
because $\dot{S} \propto \dot{T}/T^2 \propto -\dot{H}/H^3$. 

We have two important reasons why the description of 
equilibrium thermodynamics can be found. 
First, the energy density and pressure of dark components can be 
redefined and the standard continuity equation can be met. 
Second, we have $S=A/(4G)$, similarly to that in general relativity. 
We mention that the horizon entropy has been analyzed in the four-dimensional 
modified gravity~\cite{Wang:2005bi}. 
Moreover, the quantum logarithmic correction to the horizon entropy 
has been examined~\cite{Cai:2008ys, Zhu:2008cg, Lidsey:2008zq, Lidsey:2009xz, Cai:2009ua, MohseniSadjadi:2010nu}. 

The horizon entropy $S$ in the description of equilibrium thermodynamics 
is related to that $\hat{S}$ in the description of 
non-equilibrium thermodynamics as~\cite{Bamba:2009id, Bamba:2011jq} 
\begin{equation}
dS = d\hat{S} + d_i \hat{S} 
+\frac{\tilde{r}_A}{2GT_{\mathrm{H}}} dF 
-\frac{2\pi \left(1-F\right)}{G}
\frac{\dot{H}}{H^3}\, dt. 
\label{eq:3.31}
\end{equation}
With Eqs.~(\ref{eq:3.19}) and (\ref{eq:3.27}), Eq.~(\ref{eq:3.31}) 
is expressed as 
\begin{equation}
dS=\frac{1}{F} d\hat{S}+\frac{1}{F}
\frac{2H^2+\dot{H}}{4H^2+\dot{H}}\,d_i \hat{S}\,.  
\label{eq:3.32}
\end{equation}
In $f(T)$ gravity, $S$ is not equal to $\hat{S}$ because $dF \neq 0$. 
On the other hand, if $f(T)=T$, since $F=1$, we obtain $S = \hat{S}$. 
It follows from Eq.~(\ref{eq:3.32}) that 
the information of both $d\hat{S}$ and $d_i \hat{S}$ 
in the description of non-equilibrium thermodynamics 
are included in $dS$ in the description of equilibrium thermodynamics. 

In the flat FLRW background, for any gravity theory, 
the Bekenstein-Hawking entropy evolves as 
$S \propto H^{-2}$. 
Hence, when $H$ becomes small, 
$S$ becomes large, 
while $S$ decreases if $H$ increases. 
Such behaviors are also seen in superinflation. 
That is, 
$S$ becomes large when 
$w_{\mathrm{t}} \equiv P_{\mathrm{t}}/\rho_{\mathrm{t}} > -1$ 
($\dot{H}<0$), 
whereas $S$ becomes small if 
$w_{\mathrm{t}}<-1$ ($\dot{H}>0$). 
This property is similar to that in general relativity, 
in which the energy density and pressure of dark components 
are given by $\rho_{\mathrm{DE}}$ in Eq.~(\ref{eq:3.24}) and 
$P_{\mathrm{DE}}$ in Eq.~(\ref{eq:3.25}), respectively. 

The Wald entropy behaves as $\hat{S} \propto FH^{-2}$, 
where the information of gravity theories is involved. 
As an example, if $f(T)= T + \alpha T^n$ with 
$\alpha$ and $n$ constants, we find 
$\hat{S} \propto H^{2(n-2)}$. 
Accordingly, since $H$ becomes large (small) for $n>2$ ($n<2$), 
$\hat{S}$ increases apart from $n=2$. 
Hence, the evolution of the Wald entropy 
is not equivalent to that of the Bekenstein-Hawking entropy. 
A relation between the descriptions of equilibrium 
non-equilibrium thermodynamics is presented 
by the term $d_{i}\hat{S}$ of the entropy production 
in the description of non-equilibrium thermodynamics 
in Eq.~(\ref{eq:3.32}), in which 
$dS$ is constructed by $d\hat{S}$ and $d_{i}\hat{S}$. 
Thus, the description of equilibrium thermodynamics 
is clearer than that of non-equilibrium thermodynamics. 
In the description of equilibrium thermodynamics, 
the expression of the horizon entropy is 
equivalent to that in general relativity. 
Furthermore, 
the non-equilibrium thermodynamics is connected 
with the equilibrium thermodynamics more profoundly. 

We remark that the EoS of dark components in 
the description of equilibrium thermodynamics 
is different from that in 
the description of non-equilibrium thermodynamics. 
With Eqs.~(\ref{eq:3.7}), (\ref{eq:3.8}), (\ref{eq:3.24}), 
and (\ref{eq:3.25}), we obtain
\begin{eqnarray} 
\hat{w}_{\mathrm{DE}} = {\hat{P}_{\mathrm{DE}} \over \hat{\rho}_{\mathrm{DE}}}
\Eqn{=} -1 + \frac{4H\dot{F}}{FT - f}\,,
\label{eq:3.33} \\ 
w_{\mathrm{DE}} = {P_{\mathrm{DE}} \over  \rho_{\mathrm{DE}}}
\Eqn{=} -1 -\frac{4\left(1 -F -2TF^{\prime} \right) \dot{H}}
{-T -f +2TF}\,. 
\label{eq:3.34}
\end{eqnarray} 
If $f(T)$ is not equal $T$, we find 
$\hat{w}_{\mathrm{DE}} \neq w_{\mathrm{DE}}$. 
Therefore, in $f(T)$ gravity, when we compare the theoretical results 
on the EoS of dark components with the observations, 
we need the representations of the EoS of dark components in both the 
descriptions of non-equilibrium and equilibrium thermodynamics. 
This is a significant cosmological consequence acquired 
from the considerations of descriptions of non-equilibrium and equilibrium thermodynamics.

\subsubsection{Second law of thermodynamics in equilibrium description} 

The Gibbs equation for all the matter and energy fluids 
in the description of equilibrium thermodynamics is written as 
\begin{equation}
T_{\mathrm{H}} dS_{\mathrm{t}} = d\left( \rho_{\mathrm{t}} V \right) +P_{\mathrm{t}} dV 
= V d\rho_{\mathrm{t}} + \left( \rho_{\mathrm{t}} +P_{\mathrm{t}} 
\right) dV\,. 
\label{eq:3.35}
\end{equation}
We can represent the second law of thermodynamics as 
\begin{equation}
\frac{dS_{\mathrm{sum}}}{dt} 
\equiv 
\frac{dS}{dt} + \frac{dS_{\mathrm{t}}}{dt} 
\geq 0\,,
\label{eq:3.36}
\end{equation}
with $S_{\mathrm{sum}} \equiv S + S_{\mathrm{t}}$. 
{}From $V=4\pi \tilde{r}_A^3/3$, and 
Eqs.~(\ref{eq:3.13}), (\ref{eq:3.23}), and (\ref{eq:3.30}), we acquire 
\begin{equation}
\frac{dS_{\mathrm{sum}}}{dt} = 
-\frac{6\pi}{G}\left( \frac{\dot{T}}{T} \right)^2 \frac{1}{4HT + \dot{T}}\,.
\label{eq:3.37}
\end{equation}
The condition $Y \equiv -\left( 4HT + \dot{T} \right) 
= 12H \left( 2H^2 + \dot{H} \right) \geq 0$ 
follows from Eqs.~(\ref{eq:3.36}) and (\ref{eq:3.37}). 
In the flat FLRW space-time for the expanding universe, in which 
we have $H>0$, if $\left( 2H^2 + \dot{H} \right) \geq 0$, 
the second law of thermodynamics can be met. 
For $F(R)$ gravity, in the flat FLRW background, we have 
$R=6\left( 2H^2 + \dot{H} \right)$~\cite{BG-1}. 
Since $R \geq 0$, in this case, the condition $Y\geq 0$ 
shown above is satisfied. 
The condition $Y\geq 0$ can be extended to $f(T)$ gravity 
due to the analogy with $F(R)$ gravity 
because $Y$ consists of only 
$H$ and $\dot{H}$ and has a relation to the scalar curvature for general relativity. 
As a result, a unified picture of 
descriptions of non-equilibrium and equilibrium thermodynamics. 
has been realized. 
This consequence can be verified only if 
the temperature of the universe outside the apparent horizon 
is equal to that inside it~\cite{Gong:2006ma, Jamil:2009eb}.

\section{Conclusions}

In the present paper, we have reviewed the relationship between gravitation and thermodynamics. Particularly, we have considered the profound connection 
of various modified gravity theories, in which the late-time accelerated 
expansion of the universe can be realized, to thermodynamics. 

First, we have shown that the gravitational field equations in modified gravity theories of $F(R)$ gravity, 
the scalar-Gauss-Bonnet gravity, $F(\mathcal{G})$ gravity, and
the non-local gravity are equivalent to the Clausius relation in
thermodynamics. 
For the representation of the entropy in modified gravity theories, 
it is significant if the contribution from matter with or without 
the modified gravity is included to the definition of the energy flux, 
i.e., heat. 
This is relevant to the point that 
for modified gravity theories, in the expanding universe, 
a description of equilibrium thermodynamics 
on the apparent horizon can exist by redefining an energy momentum tensor of modification of gravity from which a local energy conservation is met~\cite{Bamba:2009id, Bamba:2011jq}. 

Second, for $f(T)$ gravity, we have explored 
the first and second laws of 
thermodynamics of the apparent horizon. 
We have studied both the descriptions of non-equilibrium and equilibrium thermodynamics. 
The same dual equilibrium/non-equilibrium formulation in $f(T)$ 
as in $F(R)$ gravity have been found. 
In addition, it has been verified that 
when the cosmic temperature of inside of the apparent horizon is 
the same as that of the horizon, 
the second law of thermodynamics can be met 
in the frameworks of non-equilibrium and equilibrium thermodynamics. 
We have demonstrated that 
in the description of non-equilibrium thermodynamics, 
the second law of thermodynamics can hold 
independent of sign of the time derivative in terms of 
the Hubble parameter, 
while 
in the description of equilibrium thermodynamics, 
the second law of thermodynamics can be satisfied 
thanks to the analogy with the point that 
in the expanding space-time, 
the non-negative quantity relevant to the scalar curvature 
in general relativity is positive or equal to zero. 

It is considered that 
the second law of thermodynamics in $f(T)$ gravity 
discussed in this work has a physical meaning. 
In any successful gravity theory 
alternative to general relativity, 
the second law of thermodynamics should be met. 
When the second law of thermodynamics does not hold 
in a cosmological model, 
this may comes from the fact that 
the second law of thermodynamics is not generalized 
correctly or that the model has some inherent inconsistent points. 
In the latter case, the model has to be abandoned.
Moreover, the following point should be cautioned. 
The fact that the cosmic temperature of inside of the apparent horizon 
is equal to that on the horizon 
is a working hypothesis because this cannot be so in general.

\section*{Acknowledgments}

The author would like to sincerely thank Professor Sergei D. Odintsov for very useful suggestions and kind encouragements on this research. This work was partially supported by the JSPS Grant-in-Aid for Young Scientists (B) \# 25800136 and
the research-funds presented by Fukushima University (K.B.).

\end{document}